\documentclass{elsart}

\usepackage{epsfig}

\usepackage{amssymb}

\begin{document}
\begin{frontmatter}
\title{Optimization as a result of the interplay between dynamics and 
structure}  
\author{Mateu Llas\corauthref{mat}},
\ead{mateu@ffn.ub.es}
\author{Pablo M. Gleiser},
\author{Albert D\'{\i}az-Guilera},
\author{Conrad J. P\'{e}rez}
\corauth[mat]{Corresponding author: 
 Tel:+34 93 4021150 Fax:+34 93 4021149}

\address{
        Departament de F\'{\i}sica Fonamental, Universitat de
        Barcelona, Diagonal 647, E-08028
        Barcelona, Spain \\}

\begin{abstract}
In this work we study the interplay between the dynamics of a model  of diffusion governed by a
mechanism of imitation and its underlying structure. The dynamics of the model can be quantified by
a macroscopic observable which permits the characterization of an optimal regime. We show that dynamics
  and underlying network cannot be  considered as separated ingredients in order to achieve an optimal behavior.
\end{abstract}

\end{frontmatter}

\section{Introduction}

In recent times the possibility of using the tools of statistical
physics to  analyze the rich dynamical behaviors observed in
social, technological, biological and economical systems has
attracted a lot of attention from the physics community. So far,
one of the main contributions to these fields has been the
analysis of simple models that capture the basic features of the
investigated phenomena. The goal is to identify their relevant
ingredients as well as the essential mechanisms governing their
dynamics with the hope that this information will help us to
understand the physical behavior of real complex systems. A great
part of this effort has been devoted to the characterization of
real networks, identifying their main features, and
understanding how they arise  since in nature the coupling and
interaction between units always has a clear motivation
\cite{Strogatz,Albert,Mendes,Tesfatsion}. This approach provides a valuable
tool to analyze how the underlying network affects the dynamics of
interest. This is a natural step forward along this line of
investigation.

For example, studies of an Ising model defined on a small world
network \cite{Barrat} showed that for any finite disorder strength
the system undergoes a crossover to a mean-field like region for
small temperatures in the thermodynamic limit. Also, the analysis
of  a biological evolution model defined on a small world
\cite{Bagnoli} has shown that even for a small number of quenched
long-range jumps in the genotypic space the results are
indistinguishable from those obtained by assuming all mutations
equiprobable. The dramatic change in some static quantities such
as the mean distance between units has been argued as the key
mechanism to explain these phenomena. However, there is another
set of works pointing out that other elements are necessary in
order to observe changes when the underlying structure is
modified, something which cannot be directly related to known
static properties of the network. Along this line, numerical
simulations on a non-equilibrium directed Ising model on a small
world \cite{Sanchez} showed a non trivial phase diagram. This
system exhibits a line of continuous phase transitions below a
critical threshold of disorder, above which the transition becomes
first order. In agreement to this model other systems display
transitions between different regimes but only when a finite
number of shortcuts is introduced. In a model for the spreading of
rumors \cite{Zanette} this critical value separates two regimes.
In the first  regime the rumor is bound to a finite neighborhood
of an initially infected site while on the other one it reaches a
finite fraction of the population. A similar phenomenon  was
observed in a model for the spreading of diseases \cite{Kuperman}
where a transition from an endemic situation to an oscillatory one
was found.

These are just a few modern examples where significant changes in
the properties of models are observed when their underlying
structure is modified. Our work fits in this intriguing area.
However, we will not  focus our attention only on  what happens
when a structure is replaced by another. There are systems whose
behavior can be described in terms of macroscopic observables, which  
give an idea about how well a system performs in a
given scenario. Precisely, in this work the interplay between the dynamics and
the underlying network is going to be exploited in order to reach
an optimal behavior. Keeping this idea in mind we will show that a
clear separation between both ingredients is not possible. In this
way given a particular set of features which characterize the
system, we wonder for the binomial dynamics/structure that
optimizes its performance, measured in terms of the appropriate
magnitudes. As we will see the answer is not trivial, even for the
simple model of diffusion of innovations
\cite{Guardiola2,Guardiola} that will be under consideration in
this paper.

The outline of this work is as follows. First we will define the
model, and describe its main features. Then we will analyze the
influence of different underlying networks on the dynamical
properties of the system. Using this information, we will study the
interplay between structure and dynamics. Finally, we will show
that both elements need to be taken into account in order
to achieve an optimal regime.

\section{The model}

Our starting point is a model of diffusion of innovations by
imitation recently proposed in \cite{Guardiola}. In spite of its
apparent simplicity it displays a very rich behavior
presenting self-organization, subcritical, critical and
supercritical regimes and deserves further analysis. The model 
was originally presented in a computational economics context 
\cite{Tesfatsion}, 
describes how
 an external perturbation that affects
individuals propagates to the rest of the population by a
mechanism of imitation. Such a perturbation, modeled in terms of
random updates, is independent of the underlying network and may
affect any agent in the system. When an agent changes her state
such information will be available only to the agents which are in
contact with her. These agents should decide if it is beneficial for them to
imitate, or remain in their present state. In the case that any of
 these agents decides to imitate the information
will also propagate to her neighbors. The process ends
when everybody is satisfied with their current situation. In this
way, the features that characterize the members of the population
are diffused as a wave of changes throughout the system,
eventually reaching all agents. Clearly, the main
features of the whole process will depend on how we model the
mechanism of imitation as well as the way in which the agents are
connected.

To be more precise, let us explicitly enumerate the dynamical
rules governing the behavior of the individuals. Each site (or
{\em agent}) $i$ is characterized by a real variable $a_i$.  In a
general way we can consider this quantity as a {\em
characteristic} of a given individual that other agents might want
to imitate. When an agent has adopted a new characteristic, her
neighbors become aware of the change and balance their interest
(quantified as $a_i-a_j$) with their resistance to change $C$ to
decide if they would like to imitate this change.  In this way $C$
controls the mechanism of  imitation. This parameter is constant and
the same for all the agents in the system.

The dynamics can be summarized as follows \cite{Guardiola}:
\begin{enumerate}

\item The system is asynchronously updated, that is, at each time step a
randomly selected agent $a_i$ updates her state
\begin{equation}
a_i \Rightarrow a_i + \Delta_i,
\end{equation}
where $\Delta_i$ is a random variable exponentially distributed
with mean $\lambda$. This driving process accounts for the
external pressure that may lead an individual to spontaneously
adopt a new characteristic.

\item All agents $j \in \Gamma(i)$, where $\Gamma(i)$ is the set of neighbors of
agent $i$, decide whether they want to upgrade or not, according
to the following rule
\begin{equation}
a_i -a_j \ge C \Rightarrow a_j = a_i,
\end{equation}

\item If any $j \in \Gamma(i)$ has decided to imitate, we also let her neighbors decide whether they
want to imitate this behavior or not. In this way the information of an update may spread beyond the first neighbors
of the originally perturbated site. This procedure is repeated
until no one else wants to change, concluding an {\em avalanche} of imitation events.
In this way we have assumed that the time scale of the imitation process
is much shorter than the one corresponding to the external driving.
\end{enumerate}

Regardless of the particular connectivity pattern between agents,
there are some common trends that helps us to understand
intuitively the collective behavior of the system. Hence, it is
easy to understand the role of the parameter $C$ and see how
different dynamical regimes come out in a natural way. For small
$C$ there is almost no resistance to change, and the information
spreads easily through the  system. Avalanches involve a large
number of units and therefore agents are continually updating
their state, although by small amounts. On the other hand, for
large $C$, since the resistance to change is high, the number
of units involved in the diffusion process is rather small which
implies a localized region of propagation for most of the updates.
As a consequence, distant agents can present very different
characteristics, and only in few
occasions they are able to find out about their differences. When this
happens the changes (variation of $a_i$) adopted will be large.
For intermediate values of $C$ the behavior is rather complex and
the most interesting from a collective standpoint.

A desirable situation would be one in which most of the time the
agents adopt changes that lead to large advances. In this context
the following question  arises: what is the resistance to change
that permits the agents to reach  a given average level  with  a
minimum number of upgrades? By measuring the mean rate of advance
$\rho$  it has been shown that there exists a
unique value of $C$ for which this optimal regime can be achieved \cite{Guardiola}.
This quantity is  simply defined as the ratio between  the total
advancement over the total number of upgrades, that is

\begin{equation}
\rho = \lim_{T \to \infty} \rho(T) = \lim_{T \to \infty}
\frac{\sum_{t=0}^{T} A(t)}{\sum_{t=0}^{T} s(t)}
\end{equation}

\noindent
where $s(t)$ is the number of sites that have changed (i.e.
the avalanche size) at time $t$, and $A(t)=\sum_i
\left(a_i(t)-a_i(t-1)\right)$ is the advance triggered by the
avalanche. The optimal growth regime is characterized by the
presence of a peak, $\rho_{max}$, in the mean rate of advance
 for a given value of the parameter $C$. This peak scales
with system size as a power law  $\rho_{max} \sim N^{\alpha}$.
This property is very convenient in certain frameworks. For example, from
an economic point of view the model displays scale effects, i.e., large economies grow faster.

So far, we have considered general aspects of the model. Since the
main goal of this paper is to analyze the interplay between
underlying structure and dynamics, let us put some attention in
the influence of the connectivity pattern on the collective
properties of the system and, in particular, in macroscopic
observables such as $\rho$. There are two simple cases which are
systematically analyzed: a fully connected network and a regular
lattice. In the globally coupled case the  information referent to
any change elicited in an arbitrary position of the network is
immediately available to every other agent. As a consequence, the
state of all the agents $a_i$ is bound in a gap of width $C$.
This limits the advance of any agent to a maximum of $C$. On the
other hand, when the system is defined on a 1D ring this
limitation only applies to the nearest neighbors. If the
information spreads beyond these neighbors, the advance achieved
can exceed $C$. In this extreme cases different qualitative
behaviors are observed, which can be quantified by the exponent
$\alpha$. When one considers the dynamics of the model on a ring
$\alpha = 0.20(2)$ while mean field calculations and numerical simulations of the dynamics of
the model on  a fully connected network show that $\alpha =0.50$
\cite{Guardiola}.

Summarizing, we have presented a model that  presents
an optimal  regime which can be characterized by the mean rate of
advance $\rho$. The scaling properties of this magnitude can be
used to quantitatively distinguish among different dynamical
behaviors. These dynamical behaviors have been observed in two
particular underlying networks. In the next section we will
consider more general structures and analyze their effects on the
dynamical properties of the model.

\section{Scaling analysis}

The two particular cases considered so far are usually chosen
either for their numerical simplicity, as in the ring, or because
they allow for simple mean field calculations as in the fully
connected network. Clearly, the structure of both cases is far from
the much more complex pattern of interactions observed in
realistic systems. However, they appear as paradigms of two opposite
generic situations either a scenario where the propagation of
information can be constrained  to a local neighborhood, or one where it may
reach the whole population in just one step.  One wonders which will
be the dominant features in  a more general situation. In this
section we will consider this issue. In particular, we will study
the relation  between the salient features present in a more
general structure and the dynamical properties of the system.

\begin{figure}
\label{fig1}
\centerline{
        \epsfxsize= 8.0cm
        \epsffile{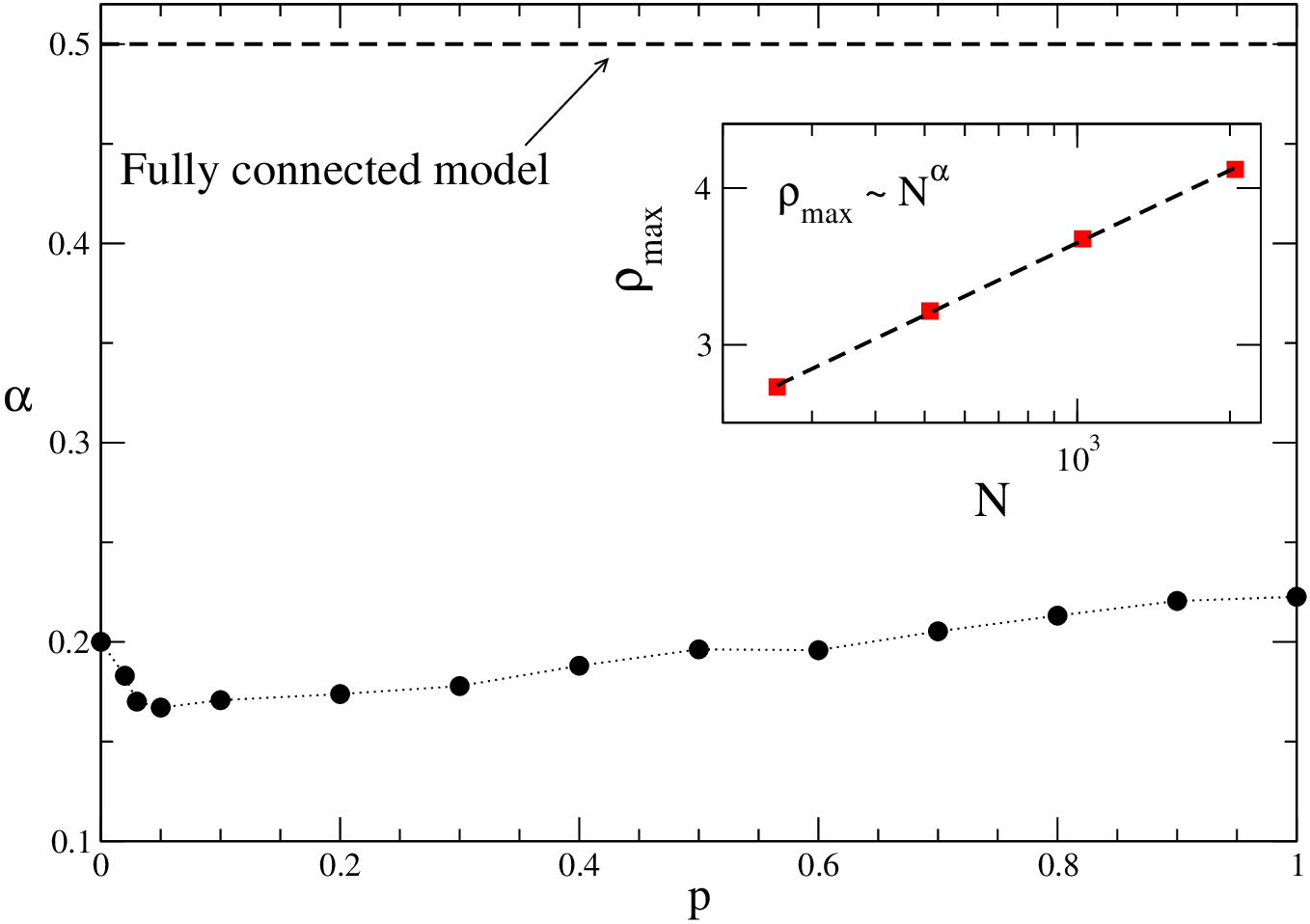}}
        \caption{The exponent $\alpha$ characterizes the  power law divergence  $\rho_{max} \sim N^{\alpha}$. 
The figure displays  $\alpha$ vs $p$. The inset shows the fit used to obtain $\alpha$ when $p=0.5$.}
\end{figure}

It is well known that, starting from a ring, the random addition
of a few links produces changes in the  properties of the network,
such as a rapid drop in the average distance between nodes,
maintaining the local structure \cite{Watts}. It has been shown
that this feature can be related to significant changes in the dynamical properties in some systems
\cite{Barrat,Bagnoli}. In order to study what effects may be
present in our model, we will consider the dynamics on a ring
lattice with $N$ vertices and $k=2$ edges per vertex,  adding a new
link at random with probability $p$ per edge \cite{Newman1}. In
general, when one modifies the underlying structure, a
quantitative variation of the numerical values of the dynamical
magnitudes that characterize the system is observed. However, we
will focus our attention on whether the scaling properties are
modified, since they are an indication of qualitative changes in
these properties. In order to quantify the changes in the scaling
behavior of the system we have computed the exponent $\alpha$ for
different values of $p$, as shown in  Fig. \ref{fig1}.

The main conclusion that can be extracted from the figure is that
the addition of new links does not modify substantially the
scaling properties of the system. For small $p$ we observe a small
decrease in the value of $\alpha$, effect that seems to be
correlated with the rapid drop in the average distance between
agents. As $p$ increases to $p=1$ a slight growth of $\alpha$ is
reported. It is important to understand why the qualitative behavior
of the system is similar to the one characteristic of the ring. In
this case, an avalanche propagates through steps in which the
information reaches the nearest neighbors of a modified site. To
generate a large event or simply to reach agents far away from the
initial updated unit a large number of steps is required. In this
sense, the information propagates by a local process. The addition of new links allows the information to reach
agents through short cuts, but it does not change the mean
mechanism of diffusion, i.e. in order to proceed further, an
avalanche still requires a large number of steps, it is still
dominated by a local process.  One cannot under-stress the fact
that for the imitation strategy described in this paper, in
contrast to what is observed in other models, the characteristic
behavior of the ring is dominant even when a more general
structure, such as a small world network, is considered.

\begin{figure}
\label{fig2}
\centerline{
        \epsfxsize= 8.0cm
        \epsffile{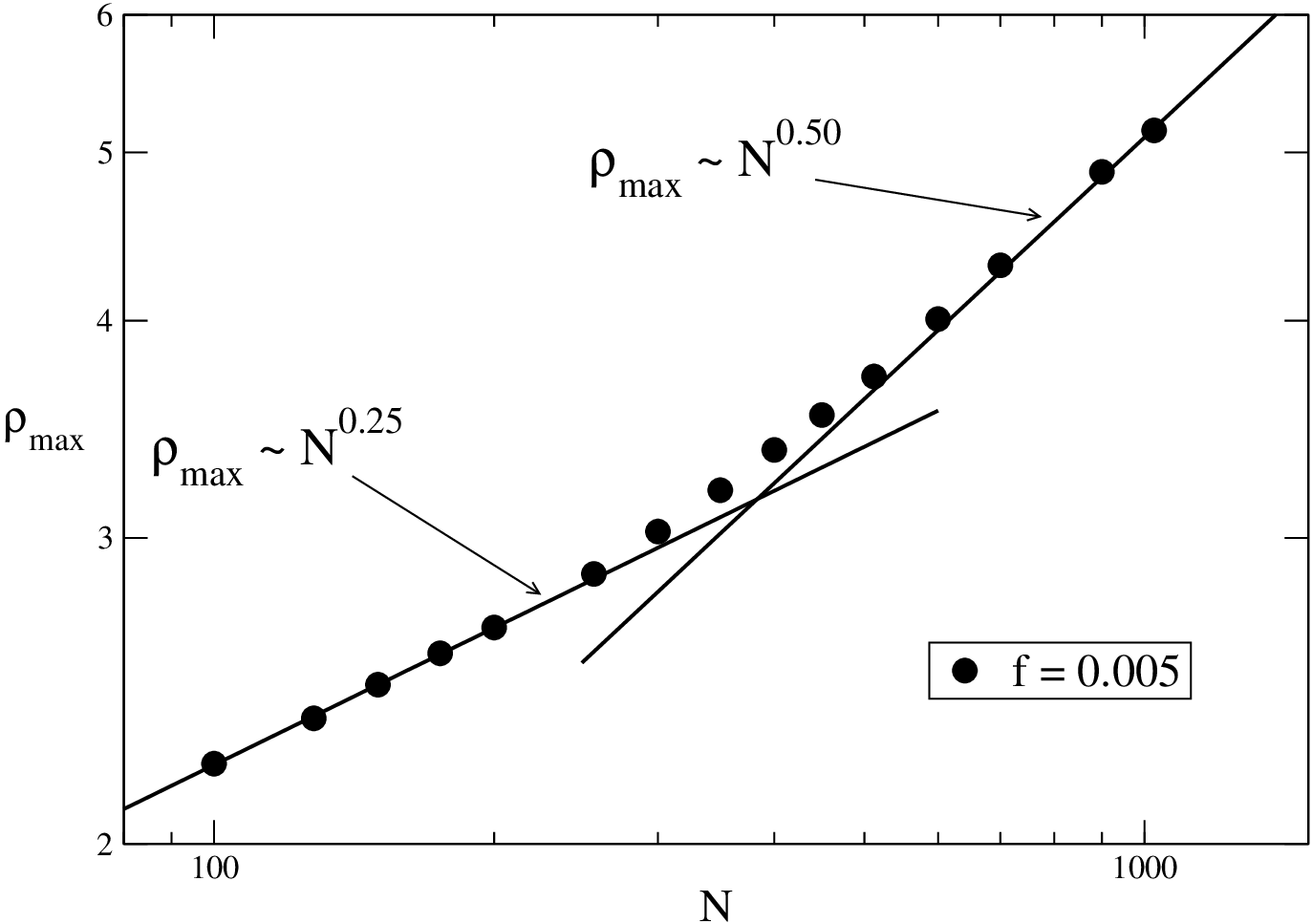}}
        \caption{The peak of the mean rate of advance $\rho_{max}$ vs $N$ for $f=0.005$. A clear 
        crossover to $\rho_{max} \sim N^{0.5}$ is observed for sufficiently large $N$. The straight 
        lines are a guide to the eye.}
\end{figure}

The local character of the diffusion process typical of
low connectivity networks is lost when considering densely
connected networks. In this situation the information about the state of a given agent is available to
any other agent in just a few steps. The small world construction
is far away from this limit, even when $p=1$ the mean number
of links per site, i. e. the connectivity of the system, has
increased very slightly. In order to reach a highly connected
state we added links randomly to the ring up to a fraction
$f$ of the total possible links in the system. This means
adding $f[N(N-1)/2]-N$ connections to the ring. This recipe allows us to
interpolate  between the ring ($f=0$)  and  the fully
connected model ($f=1$). For finite values of $f$ the
system corresponds to a ring with a superimposed random graph. For
this construction the connectivity of the system will be
$fN$, and thus, for a fixed value of $f$, the
connectivity will increase  as $N$ grows. If this
plays a significant role in the dynamical behavior of the system, 
we expect that it will be reflected in an important quantitative change in
$\alpha$. In Fig. \ref{fig2} 
we present the behavior of the peak
of the mean rate of advance $\rho_{max}$ as a function of system
size $N$ when $f=0.005$. For low values of $N$  the behavior
resembles the one observed in the small world case. As $N$ grows there is a crossover to the
fully connected behavior and $\rho_{max} \sim N^{0.5}$.

\begin{figure}
\label{fig3}
\centerline{
        \epsfxsize= 8.0cm
        \epsffile{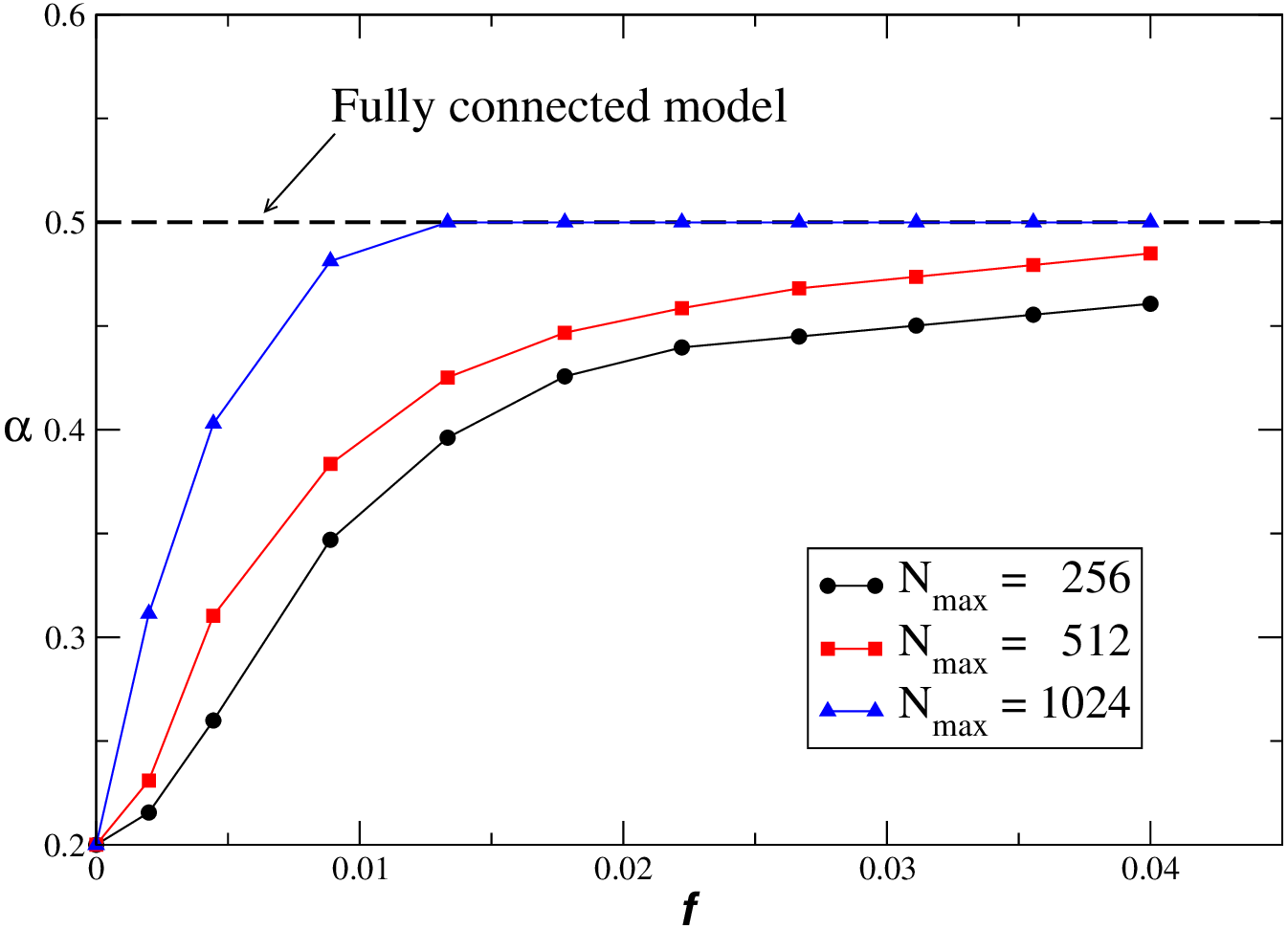}}
\caption{Exponent $\alpha$ vs
$f$. Finite size effects are clearly present. The maximum
system sizes used in the fits are $N=256$ (circles), $N = 512$
(squares) and $N = 1024$ (triangles). The lines are a guide to the
eye.}
\end{figure}

To complete the analysis we have performed a finite size study of
the crossover between both regimes. The results are presented in
Fig. \ref{fig3}. 
As one can expect, and in agreement with the
previous discussion, the jump from the ring-like scaling behavior
to the highly connected network scaling behavior becomes sharper
as N grows being the crossover located at $f
\approx \frac{1}{N}$. Numerical simulations support this result
suggesting that in the thermodynamic limit a transition to
$\alpha=0.5$ should be present for any non-vanishing value of
$f$, i.e. provided the connectivity of the system grows with
the system size. Therefore, by using this construction we have
found that there are two dominant behaviors, one for networks with
a number of links scaling with the size of the system, $O(N)$, and
another where the connectivity grows with the size of the system,
i.e., the number of links go as $O(N^2)$. It will be very interesting
to analyze the situation for other complex situations such as
scale free networks. This is currently under study and will be
published elsewhere.

We have remarked the fact that the behaviors observed in
the ring and the fully connected network were generic regarding
the way the information propagates through the system. We found
that these dynamical behaviors are dominant in more complex
structures. In a small world network the dynamics is dominated by
local processes. In order to reach the qualitative behavior
observed in a fully connected network a higher connectivity is
necessary. In fact, we have shown that by introducing a mechanism
which takes this effect into account we were able to reach this
qualitative behavior.  The identification of the dominant
behaviors of our model in more general networks is an interesting
issue in itself, but it is also an essential step in exploiting the 
interplay between dynamics and structure to reach an optimal regime. 
Precisely,  to consider both elements in a unified framework
is the main purpose of this work.  The following section is devoted to 
the investigation of our model under this new exciting point of view.

\section{Identification of an optimal behavior}

Up to now, we have analyzed different mechanisms for the spreading of the information.
On one hand, $C$ appears as a parameter that can be tuned
in order to reach an optimal regime for a fixed underlying
network. On the other hand, we have also seen that diverse
structures lead to different ways in which the information
spreads. This line of reasoning naturally leads to the question of
which is the structure that gives the optimal regime for a fixed
value of $C$. For small $C$ there is almost no resistance to
change and the information easily spreads. In fact we expect that
if $C \to 0$ then $\rho$ will also decrease independently of the
underlying network. On the other extreme, for $C \to \infty$ the
behavior of the system will resemble a random deposition process,
and again the behavior of $\rho$ is expected to decrease when
considering any general structure. To analyze what happens for
intermediate values of  $C$ we should proceed with care as we will
see immediately.

In order to do this we consider a system with a certain
distribution of couplings between agents and a fixed $C$. In
general, the value of $\rho$ will be smaller than $\rho_{max}$.
Now, suppose that we can modify the connectivity pattern in order
to reach the optimal behavior. What should we do? Is it better to
increase the number of links and tend to a densely connected
network or, on the contrary, we should go towards a regular
structure with a small number of connections per site? The
question cannot be answered properly without looking at the
parameter that control the dynamics, i.e., the strategy to follow
depends precisely on $C$. To optimize the behavior of the system
one cannot split the problem in two independent parts; dynamics
and the underlying structure must be considered as a whole.

Let us analyze some features associated to the mechanism of
imitation and consider the difference between the highest
($a_{max}$) and the lowest ($a_{min}$) characteristic values in
the system. For sufficiently high connectivity this gap will be
limited to $a_{max}-a_{min} \le C$ and, as a consequence, for
these systems the value of $\rho$ cannot exceed $C$. On the other
hand, when the propagation of the information is constrained to
advance through local processes, a more heterogeneous profile of
characteristic values can be formed, allowing for a larger gap
between $a_{max}$ and $a_{min}$. In this case, avalanche events
consisting of a large number of steps will produce a large advance,
allowing the value of $\rho$ to exceed $C$. 
In fact, by using the probability distribution of avalanches $P(s)$ and advances $P(H)$
obtained using numerical simulations in \cite{Guardiola} this can be easily verified analitically.
In this situation a sparsely connected network will necessarily have a greater $\rho$
than a highly connected one. For increasing values of $C$,
avalanche events will easily get blocked in a few steps. In this
context, large advances in a sparsely connected network will
become very rare. More frequent advances will be observed in a
highly connected network since many agents are permitted to find
out about an update in any step. This situation offers the
possibility for a higher $\rho$ to be observed in highly connected
structures.

\begin{figure}
\label{fig4}
\centerline{
        \epsfxsize= 8.0cm
        \epsffile{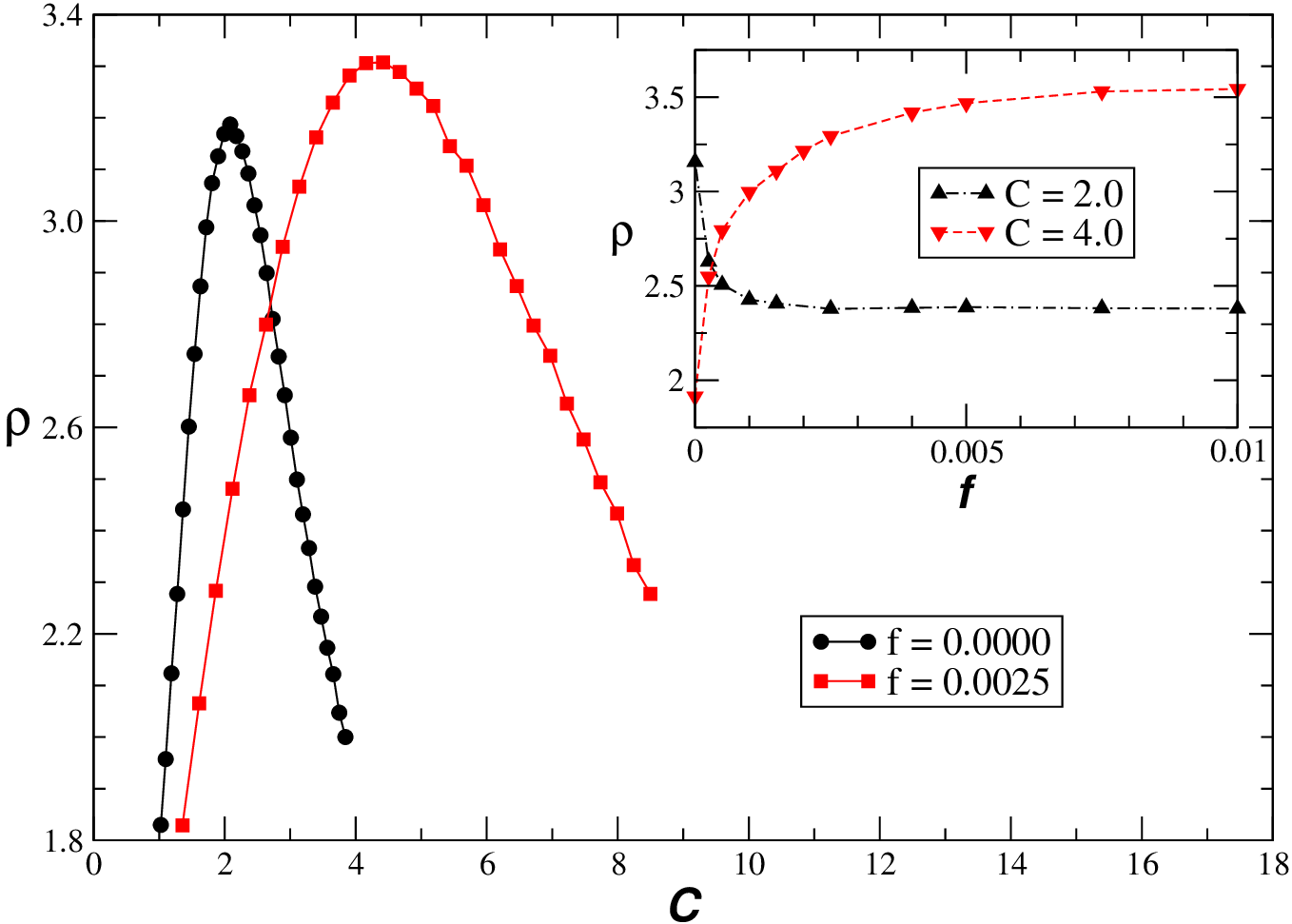}}
 \caption{Mean rate of advance
$\rho$ vs. $C$ for a fixed system size $N=512$ and for
$f=0.00$ and $f=0.0025$. The inset presents the behavior
of $\rho$ vs. $f$ for $C=2.0$ and $C=4.0$.}
\end{figure}

Following this analysis we have considered a general
system of fixed size $N$. The evolution of the 
mean rate of advance $\rho$ vs $C$ has been studied for two different situations:
the ring and another structure where the number of links (measured
in terms of $f$) is large enough to observe fully connected
behavior. The results are illustrated in Fig. \ref{fig4}. 
Note that as $f$ is varied the qualitative shape of the $\rho$
curve is similar. However, the position of the peak corresponds to
different values of $C$. For increasing values of $f$ the
peak corresponds to larger values of $C$, eventually reaching the
curve corresponding to the globally coupled case. It is important
to stress that, as $C$ is varied, the optimal network may change
from a highly connected to a sparsely connected one. This behavior
is clearly reflected in the inset, where we present the behavior
of $\rho$ vs. $f$ for two different values of $C$. For $C=2$
a decrease in $\rho$ is observed as $f$ grows. The addition
of links is harmful for the system. On the other hand, for $C=4.0$
the opposite behavior is observed, and $\rho$ increases its value
as $f$ grows. Clearly, in this case,  the  addition of  links
is beneficial.

These results show that in order to optimize the behavior of the
system a non trivial combination of  dynamical rules and
underlying structure should be considered. When the interplay
between both allows for $\rho/C > 1$, a  sparsely connected
structure performs better than a highly connected one.  On the
other hand, when $\rho/C < 1$ the opposite is true. Note that when
$\rho /C \sim 1$ the behavior should be  independent of the
underlying  network. In fact, Fig \ref{fig4} 
shows that when
$\rho/C \sim 1$  both curves intersect. Numerical simulations for
different $N$ and $f$ also show these qualitative behaviors.

\section{Conclusions}

The search for  the essential mechanisms that govern the dynamics
of real complex systems has led  recent efforts to  highlight the
role of the complex underlying structures present in these
systems. A large amount of work has been devoted to the
characterization and identification of their main features and the
role they play on the dynamical properties. This scenario suggests
that a more general framework, devised to address these issues,
should consider the interplay between the dynamics and the
underlying structure as a single ingredient.

The main aim of  this work has been to take a first step into this
direction. With this picture in mind we analyzed a model of
diffusion by imitation which presents a very interesting property,
the possibility of defining an optimal regime by tuning a
parameter. This regime can be quantified by a macroscopic
observable. In order to be able to consider the interplay between
dynamics and structure we have studied the behavior of the model
on general networks and characterized the  dominant behaviors in
these structures.  An analysis of how the optimal regime can be
achieved has led to the conclusion that the dynamical rules and
the underlying structure cannot be considered as separate
ingredients.

The rich dynamical behaviors observed in the more general framework makes a detailed
analysis of the model necessary.  In particular, we are focusing our attention on
the behavior when $f \to 0$, in order to characterize a possible transition.
Since the connectivity plays an important role in the properties of the system, a
natural extension to more  general structures, such as scale-free networks, is under
consideration. Finally, a very interesting new point of view which we are also analyzing
concerns the behavior of the model on dynamical networks. Although we cannot underestimate
the results that will follow from these studies we believe that the main results presented
in this work will not be affected in their generality.



\section{Acknowledgements}
We acknowledge financial support from MCYT, grant number BFM2000-0626, 
and also from European Commission, Fet Open Project COSIN IST-2001-33555.
P. M. Gleiser acknowledges financial support from 
Fundaci\'on Antorchas.

\end{document}